\documentclass[12pt]{article}

\input{epsf.tex}

\def\IC{\bf C}
\def\IZ{\bf Z}

\def\z2z2{$\IC^3/(\IZ_2\times\IZ_2)$}

\def\beq{\begin{equation}}
\def\eeq{\end{equation}}
\def\beqa{\begin{eqnarray}}
\def\eeqa{\end{eqnarray}}
\def\tr{{\rm tr \,}}

\begin{document}
\setcounter{page}{1}

\vspace{-1cm}
\rightline{CERN-TH/2000-219, \tt hep-th/0007173}
\vspace{1.3cm}
\begin{center}
\Large{\bf From quiver diagrams to particle physics\footnote{Invited
lecture delivered at the Third European Congress of Mathematics,
Barcelona, Spain, 10-14 July, 2000}} 

\bigskip

\large{Angel~M.~Uranga \\
Theory Division, CERN \\
CH-1211 Geneva 23, Switzerland \\
Angel.Uranga@cern.ch}
\end{center}

\vspace{1cm}

\begin{center}

{\bf Abstract}
\end{center}

Recent scenarios of phenomenologically realistic string compactifications
involve the existence of gauge sectors localized on D-branes at singular
points of Calabi-Yau threefolds. The spectrum and interactions in these
gauge sectors are determined by the local geometry of the singularity, and
can be encoded in quiver diagrams. We discuss the physical models arising
for the simplest case of orbifold singularities, and generalize to
non-orbifold singularities and orientifold singularities. Finally we show
that relatively simple singularities lead to gauge sectors surprisingly
close to the standard model of elementary particles.

\newpage

\section{Introduction}

The construction of string theory models which at low energies reproduce
the basic features of the standard model of elementary particles (or some
extension thereof) is a non-trivial task from the physical point of view.
Interestingly, it also involves beautiful mathematics, and leads to a rich
interplay between the physical and mathematical points of view. To quote
a traditional example, the construction of four-dimensional ${\mathcal
N}=1$  supersymmetric heterotic string vacua \cite{chsw} involves the
study of stable $G$-bundles on Calabi-Yau threefolds, with $G$ a subgroup
of $E_8\times E_8$ or $SO(32)$ (for recent results on this approach, 
see \cite{burt}).

Recent developments in string theory have led to new kinds of string theory
vacua with potential phenomenological interest. A particular class
\cite{add} corresponds to compactification on a Calabi-Yau threefold $X$
with gauge bundles defined on subvarieties of the internal space (times
non-compact spacetime). In physical terms, the compactification includes a
set of D-branes \cite{tasi}, which are dynamical extended objects
partially wrapped on the internal manifold, and filling non-compact
spacetime. Their dynamics is controlled by a gauge theory defined on their
world-volume. The rank of the gauge bundle they carry is usually referred
to as the number of D-branes. In the following we center on the simplest
case of D3-branes in type IIB string theory, where each D-brane spans
four-dimensional spacetime times a point $P\in X$. The properties of the
corresponding gauge theory sector in the low-energy theory are then
determined in terms of the local geometry of $X$ around $P$.

For a set of D3-branes sitting at a smooth point in $X$, the low-energy
gauge field theory on the world-volume has ${\mathcal N}=4$ supersymmetry,
and is therefore non-chiral and phenomenologically uninteresting.
Chiral gauge sectors arise when D3-branes sit at singular points
in $X$. Singularities preserving at least ${\mathcal N}=1$ supersymmetry,
and at a finite distance in Calabi-Yau moduli space must be Gorenstein
canonical singularities.

In Section 2 we discuss several systems of branes at singular points in
non-compact Calabi-Yau spaces. In section 2.1 we center on the simple case
of threefold quotient singularities, which leads to an implementation of
the McKay correspondence in string theory. In section 2.2. and 2.3 we
point out several generalizations suggested by string theory. In section 3
we discuss how an extremely simple $\IZ_3$ quotient singularity leads to 
gauge groups and particle contents remarkably similar to those of the
(minimal supersymmetric) standard model.

I am grateful to G.~Aldazabal, A.~Hanany, L.~E.~Ib\'a\~nez, J.~Park,
F.~Quevedo and R.~Rabad\'an for collaboration and useful discussion on
these issues. I also thank M.~Gonz\'alez for encouragement and support.

\section{Branes at singularities}

\subsection{Branes at orbifold singularities}

Let $\Gamma$ be a discrete group of $SU(3)$, and $\{ {\bf r}_i \}$ the set
of its unitary irreducible representations. We want to consider a set of
D3-branes at the origin of ${\bf C^3}/\Gamma$, where $\Gamma$ acts on
${\bf C}^3$ through a three-dimensional representation ${\mathcal R}_{\bf
3}$.

Consider first a set of $N$ D3-branes at the origin in ${\bf C}^3$, labeled 
by an index $a=1,\ldots, N$ referred to as Chan-Paton index. Quantization
of open strings with endpoints of the D3-branes leads to a set of dynamical
fields propagating on the D3-brane world-volume. In terms of ${\mathcal
N}=1$ supersymmetry multiplets, the corresponding gauge field
theory contains a set of vector multiplets (each containing one gauge
field and one complex fermion) with gauge group $U(N)$, and three chiral
multiplets $\Phi^a$, $a=1,2,3$ (each containing one complex scalar and
one complex fermion). The latter transform in the adjoint representation
of $U(N)$ and form a triplet under the $SU(3)$ action on $\IC^3$.
The interactions are encoded in the superpotential function
$W=\epsilon_{abc} {\rm tr\,} (\Phi^a \Phi^b \Phi^c)$, where
$\epsilon_{abc}$ correspond to the components of the $SU(3)$ invariant
tensor.

Following \cite{dm} (see \cite{jm,dgm} for generalizations, and
\cite{ks,lnv,hanur,nonab} for related discussions) the field
theory on D3-branes at the origin in ${\bf C}^3/\Gamma$ is obtained from
the above field theory associated to $N$ D3-branes in flat space, by
keeping the states which are invariant under the combined action of
$\Gamma$ on ${\bf C}^3$ (as determined by ${\mathcal R}_{\bf 3}$) and on
the space of Chan-Paton indices (through a $N$-dimensional representation
${\mathcal R}$, with decomposition ${\mathcal R}=\bigoplus_i N_i {\bf
r}_i$). Following \cite{lnv}, we regard fields in the adjoint of $U(N)$ as
${\rm Hom}({\bf C}^N,{\bf C}^N)$. The projection on the ${\mathcal N}=1$
vector multiplets leaves the following fields
\beqa
{\rm Hom}({\bf C}^N,{\bf C}^N)^{\Gamma} = \bigoplus_i {\rm Hom}({\bf
C}^{N_i},{\bf C}^{N_i})
\eeqa
corresponding to a gauge group $\prod_i U(N_i)$. The projection on the
$SU(3)$ triplet of ${\mathcal N}=1$ chiral supermultiplets leaves the
following fields
\beqa
({\mathcal R}_{\bf 3} \otimes {\rm Hom}({\bf C}^N,{\bf C}^N))^{\Gamma} =
\bigoplus_{i,j} a^{\bf 3}_{ij}{\rm Hom}({\bf C}^{N_i},{\bf C}^{N_j})
\eeqa
where $a^{\bf 3}_{ij}$ are defined by ${\mathcal R}_{\bf 3}\otimes {\bf
r}_i=\bigoplus_j a^{\bf 3}_{ij} {\bf r}_j$. Hence we obtain $a^{\bf 3}_{ij}$
${\mathcal N}=1$ chiral multiplets transforming in the representation
$(N_i,{\overline N}_j)$ of the gauge group. The superpotential is obtained
by restricting the above one to the surviving fields. 

The field content on the D3-brane world-volume can be encoded in a quiver
diagram, where the $i^{th}$ node represents the $U(N_i)$ factor in the
gauge group, and $a^{\bf 3}_{ij}$ oriented arrows from the $i^{th}$ to the
$j^{th}$ node correspond to the ${\mathcal N}=1$ chiral multiplets in the
$(N_i,{\overline N}_j)$ representation. Finally, closed triangles of
oriented arrows are associated to superpotential couplings of the
corresponding chiral multiplets. The quiver for a $\IC^3/\IZ_5$
singularity is depicted in Fig.~\ref{quiver1}a.

Several interesting mathematical connections arise at this point. For
instance, the quiver diagrams encoding the field theory content and
interactions coincide with the McKay quivers associated to the singularity, 
and which are related to the homology of the resolved space by the McKay
correspondence \cite{mckay}. The correspondence arises in the string
theory context since branes giving rise to a specific gauge factor have
the geometrical interpretation of higher-dimensional branes wrapped on
homology cycles of the space (see \cite{jaume} for a detailed
description). Hence, gauge theory data (the quiver diagram) are related
to the homology of the ambient space.

Also, if ${\mathcal R}$ is chosen to be the regular representation of
$\Gamma$, the moduli space of vacua of the gauge theory corresponds to
the space of possible locations of the D3-brane, which is isomorphic to
the transverse space ${\bf C}^3/\Gamma$. The construction of the moduli
space amounts to performing a symplectic quotient in the subspace of
fields subject to relations $\frac {\partial W}{\partial \Phi_i}=0$
(F-term constraints) determined by the superpotential \cite{dgm}. It
provides the string theory counterpart of  the construction of
$\IC^3/\Gamma$ as the moduli of representations of a quiver diagram with
relations \cite{sardo}. In the particular case $\gamma\subset SU(2)$,
studied in \cite{dm,jm}, one recovers the hyperk\"ahler quotient
construction of ALE spaces \cite{kronheimer}. String theory also provides a
description of the resolved spaces, by a suitable modification of the
symplectic quotient due to a non-zero moment map (D-term).

We conclude this section with some physical considerations. In string
theory D-branes are sources of certain $p$-form gauge fields from the
closed string sector, whose equations of motion may impose certain
consistency conditions on the D-brane configuration. In the particular
case of D3-branes at ${\bf C}^3/\Gamma$ singularities, the equations of
motion for fields located at the singularity impose the so-called
twisted tadpole cancellation conditions, which for quotient singularities
amount to the vanishing of the character of the representation ${\mathcal
R}$. This constraint is equivalent to the cancellation of non-abelian
anomalies in the gauge field theory on the D3-branes world-volume
\cite{lr}. They also imply that the remaining mixed $U(1)$- non-abelian
anomalies have a factorized form and are cancelled by a version of the
Green-Schwarz mechanism \cite{iru}. The anomalous $U(1)$ factors become
massive and disappear from the low-energy dynamics.

\subsection{Generalizations}

\subsubsection{Non-orbifold singularities}

There is no simple recipe to obtain the field theory on the world-volume
of stacks of D3-branes at a general singularity. However, the 
requirement that its moduli space should correspond to the space of
possible locations of D-branes in the transverse space is enough to
determine the field theory in the simple example of the conifold
singularity $X_{con}$ \cite{kw}, defined by the hypersurface 
$x^2+y^2+z^2+w^2=0$ in ${\bf C}^4$. A set of $N$ D3-branes at a conifold
singularity yields a ${\mathcal N}=1$ supersymmetric field theory with 
gauge group $U(N)\times U(N)$ and chiral multiplets $A_i$, $B_i$, $i=1,2$
in the representations $(N,{\overline N})$ and $({\overline N},N)$ under
the gauge group, and in the representations $(2,1)_{1/2}$, $(1,2)_{1/2}$
under the $SU(2)^2\times U(1)$ symmetry group of $X_{con}$. The
interactions are determined by a quartic superpotential $W= \epsilon^{ij}
\epsilon^{kl} {\rm tr}\, A_iB_jA_kB_l$. 

This example provides a whole new family of models \cite{uraconi} (see
\cite{keshav,aklm} for a related discussion) corresponding to
D3-branes at quotients of the conifold, $X_{con}/\Gamma$, with $\Gamma$ a
subgroup of $SU(2)\times SU(2)$ in order to preserve ${\mathcal N}=1$
supersymmetry. The strategy is, as in section 2.1, to start with D3-branes
at $X_{con}$, embed the action of $\Gamma$ on the Chan-Paton indices, and
keep only fields which are invariant under the combined geometrical
and Chan-Paton action of $\Gamma$. The resulting field theories can be
encoded in a quiver diagram, with nodes and arrows corresponding to gauge
factors and chiral multiplets, and superpotential terms correspond to
closed polygons formed by four arrows. When $\Gamma$
acts on the Chan-Paton indices in the regular representation, the moduli
space of the gauge theory is a symplectic quotient subject to F-term
constraints from the superpotential, as obtained in explicit examples
\cite{unge,toriconi}, generalizing the orbifold result. The string
theory construction also suggest these quiver diagrams also encode the
homology of the resolution of $X_{con}/\Gamma$.

\begin{figure}
\begin{center}
\centering
\epsfysize=5cm
\leavevmode
\epsfbox{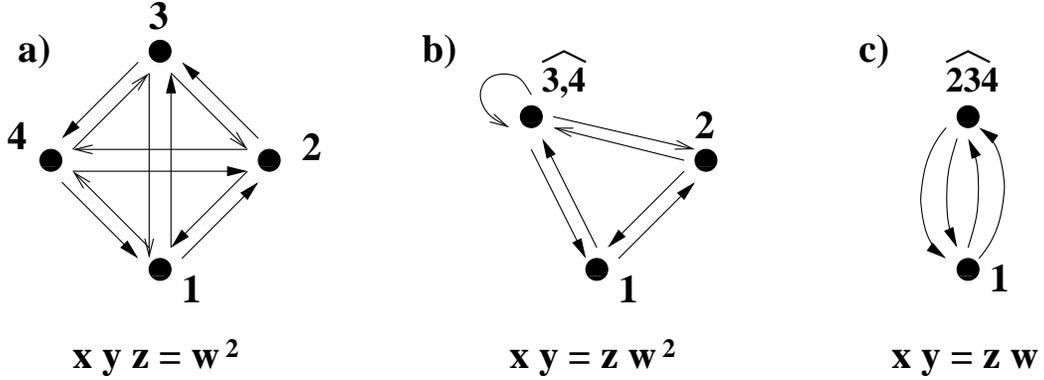}
\end{center}
\caption[]{\small 
Figure a) shows the quiver diagram for a $\IC^3/(\IZ_2\times \IZ_2)$
singularity, which can be defined as the hypersurface $xyz=w^2$ in
$\IC^4$. Partial blow-ups resolve it to the suspended pinch point
singularity ($xy=zw^2$), and the conifold ($xy=zw$), whose quiver diagrams
are shown in Figures b) and c). Each blow-up is reflected on the quiver as
the joining of two nodes (the joint node is denoted by a hat), and the
disappearance of certain arrows (depicted with thin heads).}
\label{quiver3}
\end{figure}

A general recipe to obtain the field theory on D3-branes at a general
toric singularities $X$ was proposed in \cite{mp} (see \cite{bhh} for a
more general discussion). It is based on realizing $X$ as a partial
resolution of a threefold quotient singularity ${\bf C}^3/\Gamma$ for
suitable $\Gamma\subset SU(3)$. The construction requires a precise
identification of the effect of the resolution on the field theory, which
is a specific Higgs mechanism in which several gauge factors break to
their diagonal subgroup, and some chiral multiplets become massive and
disappear from the light spectrum. The detailed map has been worked out in
some explicit examples, {\em e.g.} in \cite{mp,greene}, and provides an
explanation for the existence of quiver diagrams for non-orbifold
singularities. They are obtained by joining nodes and deleting arrows in
the quiver of the initial quotient singularity, as dictated by the Higgs
mechanism in the field theory. The quiver version of the resolution of the
$\IC^3/(\IZ_2\times \IZ_2)$ singularity to the conifold is shown in
Figure~\ref{quiver3} \cite{mp}. It would be interesting to obtain a more
precise characterization of these operations on quiver diagrams. 

Another interesting question is related to the uniqueness of the field
theory corresponding to a given singularity \cite{bhh}. In some cases, 
different field theories may have isomorphic moduli spaces. Mathematically, 
the representation moduli of the corresponding quivers with relations are
isomorphic. From the physical point of view, the equality of moduli
spaces may in some cases \cite{uraconi,unge} be a reflection of Seiberg
duality, a non-trivial infrared equivalence of seemingly different field
theories.

\subsubsection{Orientifold projections}

\begin{figure}
\begin{center}
\centering
\epsfysize=4cm
\leavevmode
\epsfbox{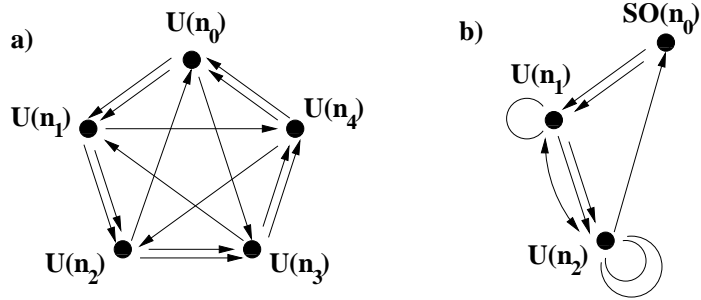}
\end{center}
\caption[]{\small Figure a) shows the quiver diagram for the $\IC^3/\IZ_5$
singularity, with $\IZ_5$ action $(z_1,z_2,z_3)\to (e^{2\pi i\frac 15}
z_1, e^{2\pi i \frac 15} z_2, e^{2\pi i\frac{-2}{5}}z_3)$. Figure b)
shows the quiver of the singularity after an orientifold projection.}
\label{quiver1}
\end{figure}

Type IIB string theory is invariant under an operation $\Omega$ which
reverses the orientation on the world-sheet. Hence it is possible to
consider modding out type IIB configurations by $\Omega$, possibly
accompanied by a geometric involution $g$, also leaving the theory
invariant \cite{orient,modernorient}. The new configurations thus obtained
are called orientifolds, and are characterized by the inclusion of
non-orientable world-sheets in the string theory perturbative expansion.

We are interested in studying D3-branes at orientifold threefold 
singularities, {\em i.e.} $\Omega g$ quotients of the system of D3-branes
at singularities in CY threefolds. The best studied case is again that of
quotient singularities ${\bf C}^3/\Gamma$, and for the sake of clarity
here we center on $\Gamma={\bf Z_N}$ with the action of the generator
$\theta$ of ${\bf Z}_N$ on ${\bf C}^3$ represented by $\gamma={\rm diag}
(e^{2\pi i t_1/N}, e^{2\pi i t_2/N}, e^{2\pi i t_3/N})$, with $t_i$
integers defined modulo $N$, and $\sum_{a=1}^3 t_a=0\,\,{\rm mod}\,\, N$.
Let us embed the action of $\theta$ on the Chan-Paton indices by a
diagonal matrix $\gamma_{\theta,3}$ with $N_k$ entries $e^{2\pi i \,
k/N}$. The field theory one obtains has vector multiplets with gauge group
$\prod_{i=1}^N U(N_i)$, and chiral multiplets $\Phi^a_{i,i+t_a}$,
$a=1,\ldots,3$, $i=1,\ldots, N$ in the representation $(N_i,{\overline
N}_{i+t_a})$.

The orientifold action may also be embedded in the space of Chan-Paton
indices, through a matrix $\gamma_{\Omega g,3}$. The representations of
$\Gamma$ and $\Omega g$ are usually constrained from mutual consistency
requirements \cite{orient,modernorient}. Several solutions to these
conditions are known (quite exhaustively in the two-fold case 
\cite{orientwo}), but a complete classification is lacking. In the
following we center on a concrete example \cite{iru} where $g$ inverts all
coordinates in ${\bf C}^3$, and exchanges closed string fields in
oppositely twisted sectors, {\em i.e.} has a non-trivial action on the
homology cycles shrunk at the singularity. We also choose $\gamma_{\Omega
g,3}$ symmetric and such that exchanges the eigenspaces of conjugate
eigenvalues in $\gamma_{\theta,3}$. The action of ${\bf Z}_N$ on
Chan-Paton indices is therefore constrained to form a real representation,
$N_k=N_{-k}$.

The effect of the orientifold projection on the spectrum is as follows.
Gauge factors associated to conjugate irreducible representations of
${\bf Z}_N$ are identified, and unitary gauge factors associated to
real representations are reduced to their maximal orthogonal subgroups.
Correspondingly, the chiral multiplets $\Phi^a_{i,i+t_a}$ and  
$\Phi^a_{-i-t_a,-i}$ are identified. When $i+t_a=-i$ the bifundamental
field is projected down to a two-index antisymmetric representation of the
unitary gauge factor after the projection. The quiver diagram for an
orientifold of a $\IC^3/\IZ_5$ singularity is depicted in
Figure~\ref{quiver1}b.

The effect on the quiver diagram of ${\bf C}^3/\Gamma$ is an identification 
of nodes and arrows related by a ${\bf Z}_2$ action, with a specific 
prescription for nodes and arrows which are mapped to themselves (it would
be interesting to characterize these mappings more precisely in order to
allow the classification of resulting models). One may wonder about the
meaning of the resulting quiver. String theory suggests it encodes the
information about the homology of the resolved orientifold singularity. In
fact, the string theory counting of homology cycles in the resolved space
(by counting of twisted sector blow-up moduli) gives roughly speaking half
the number encountered before the orientifold projection, due to the
non-trivial action of $g$ by exchanging oppositely twisted sectors. This
agrees with the counting of nodes in the orientifolded quiver, which also
gives half the number encountered before the orientifold projection.

We conclude by pointing out that orientifolds of non-orbifold
singularities  can be constructed as partial resolutions of orientifold of
quotient singularities. Preliminary results on some simple examples
\cite{pru} indicate that the effect of the orientifold projection on the
quiver of the non-orbifold singularity is also a $\IZ_2$ involution, and
that the orientifold singularity can be constructed as a symplectic
quotient on the space of fields constrained by the F-terms conditions.

\section {Particle physics}

In this section we discuss particular examples of singularities leading to
interesting low-energy field theories, in that they resemble the structure
of the (minimal supersymmetric) standard model (or some extension thereof)
which we now review. Such models would provide phenomenologically
interesting string theory vacua when embedded in a compact Calabi-Yau
context.

\subsection{Review of the (minimal supersymmetric) standard model}

All known non-gravitational interactions between elementary particles are
described by the quantum field theory known as standard model. The
simplest supersymmetric extension of this model contains vector multiplets
with gauge group $SU(3)\times SU(2)\times U(1)$. It also contains a set of
chiral multiplets transforming in three copies of the representation
\beqa
(3,2)_{1/6}+({\overline 3},1)_{-2/3}+({\overline3},1)_{1/3}+(1,2)_{1/2}
+(1,1)_{1},
\eeqa
where subscripts denote $U(1)$ charges.
Successful breaking of the electroweak interactions requires also at least
one chiral multiplet in the representation $(1,2)_{1/2}+(1,2)_{-1/2}$.

Interactions are encoded in a set of complicated gauge invariant functions
of these chiral multiplets, the superpotential and gauge kinetic
functions, which are holomorphic, and the K\"ahler potential, which is
real. We will skip their details since realistic phenomenology
usually involves additional model-dependent assumptions, like additional 
global symmetries. 

\subsection{Realistic models from $\IZ_3$ singularities}

The replication of families is an intriguing feature of the above field
theory. To reproduce it from branes at singularities, the case of ${\bf
C}^3/{\bf Z}_3$ with the ${\bf Z}_3$ action defined by $(z_1,z_2,z_3)\to
(e^{2\pi i/3} z_1, e^{2\pi i/3 z_2}, e^{2\pi i/3} z_3)$, is singled out,
in that it leads to natural triplication. The two examples we are to
consider are based on this singularity.

The first example we consider forms a subsector of the model considered in
\cite{lpt}. It is constructed by placing eleven D3-branes on the
orientifold of the ${\bf C}^3/{\bf Z}_3$ singularity, introduced in section
2.2.2. We choose $\gamma_{\theta,3}={\rm diag}(1,e^{2\pi i/3}{\bf 1}_5,
e^{-2\pi i/3} {\bf 1}_5)$. Following our rules above, we obtain vector
multiplets with gauge group $SU(5)$ (the $U(1)$ factor disappears from the
light spectrum as explained in section 2.1), and there are chiral multiplets 
in three copies of the representation $5+{\overline {10}}$. This spectrum
resembles the structure of $SU(5)$ grand unified theories, which reproduce 
the spectrum of the minimal supersymmetric standard model when an additional 
field in the adjoint representation is present to break $SU(5)$ to the
standard model group through the Higgs mechanism (an additional
$5+{\overline 5}$ pair is further required for electroweak symmetry
breaking). Unfortunately, all Higgs fields are absent in our field theory,
which therefore is suggestive but not truly realistic.

Our second example is based on the ${\bf C}^3/{\bf Z}_3$ orbifold (rather
than orientifold) singularity \footnote{This model has been studied
in\cite{aiqu} and is similar to a subsector of models in \cite{aiq1}.}, 
with Chan-Paton embedding $\gamma_{\theta,3}={\rm diag}({\bf 1_3}, e^{2\pi
i/3} {\bf 1}_2,$ $ e^{-2\pi i/3})$. This embedding does not satisfy the
tadpole cancellation condition 
stated in section 2.1 (the character of the representation of ${\bf Z}_3$ 
is non-zero). The problem can be solved by introducing an additional set
of D-branes, for instance D7$_a$-branes, with $a=1,2,3$, wrapped on the
complex surfaces $z_a=0$, which do not break further supersymmetries. For 
non-compact spaces these additional branes are non-dynamical, but they
contribute additional fields in the four-dimensional world-volume of the
D3-branes, arising from open strings stretched between D3- and D7-branes.
Hence the corresponding gauge field theories correspond to extended
quivers, where additional nodes correspond to the global symmetries
(symmetries on the D7-branes), and additional arrows correspond to 3-7 and
7-3 string states. These quiver diagrams generalize to the threefold case
those in \cite{dm}. Their moduli spaces therefore provide a generalization
of the Kronheimer-Nakajima construction of the moduli space of instantons
on ALE spaces \cite{kn}.

\begin{figure}
\begin{center}
\centering
\epsfysize=5cm
\leavevmode
\epsfbox{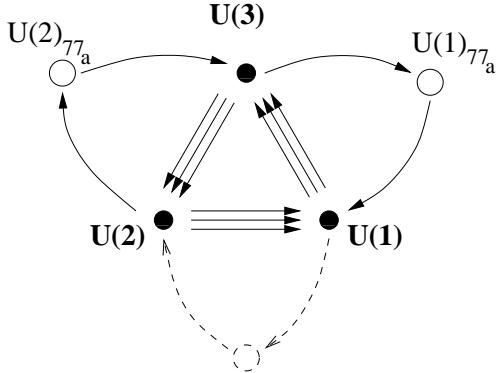}
\end{center}
\caption[]{\small Quiver diagram of a system of D3- and D7-branes at a
$\IC^3/\IZ_3$ singularity, reproducing a spectrum close to the (minimal
supersymmetric) standard model. The 3-7 sector is triplicated due to
the three kinds of D7-branes we have introduced. The dotted node and
arrows are present in the generic quiver, but not for our specific choice
of Chan-Paton representations. Notice that white nodes correspond to
global symmetries of the D3-brane field theory, and that only one of the
three $U(1)$ symmetries of the D3-branes survives at low energies.} 
\label{quiver2}
\end{figure}

The ${\bf Z}_3$ acts on the space of D7-brane Chan-Paton indices, through
matrices $\gamma_{\theta,7_a}$. The modified tadpole cancellation
conditions (which reduce to $\sum_a\tr\gamma_{\theta,7_a}+
3\tr\gamma_{\theta,3}=0$) are satisfied for the very symmetric choice
${\rm tr\,}\gamma_{\theta,7_a}=-{\rm tr\,}\gamma_{\theta,3}$, 
$a=1,2,3$. Choosing {\em e.g.} $\gamma_{\theta,7_a}={\rm diag}(e^{2\pi
i/3}, e^{-2\pi i/3} {\bf 1}_2)$ we obtain the following spectrum
\begin{eqnarray}
\begin{array}{ccc}
{\rm 3-3\,\, strings} & {\rm Gauge \,\, group} & SU(3) \times SU(2)
\times U(1) \\
& {\rm Chiral\,\, multiplets} & 3\,[\,(3,2)_{1/6} + (1,2)_{1/2} +
({\overline 3},1)_{-2/3}\,] \\
{\rm 3-7}_a {\rm \,\, strings} & {\rm Chiral\,\, multiplets} & 
(3,1)_{-1/3} + 2(1,2)_{-1/2} + \\
a=1,2,3 & & +2({\overline 3},1)_{1/3} + (1,1)_{1}  
\end{array}
\end{eqnarray}
We have included the charges under the only non-anomalous linear
combinations of the $U(1)$ factors in the original $U(3)\times U(2)
\times U(1)$ gauge group. This $U(1)$ does not become massive and plays
the role of hypercharge in the standard model we have just constructed.
The quiver diagram of this gauge theory is shown in Figure~\ref{quiver2}.

Notice how close to the spectrum in section 3.1 one can get using very
simple singularities. In particular, the later model differs from the MSSM
just in that it produces three Higgs pairs instead of one, and in fact
constitutes one of the simplest semi-realistic string models ever built.
Since such constructions would provide a rationale for the existence of
three generations (one per complex transverse dimensions), and
for hypercharge assignments (see \cite{westakt} for discussion of
hypercharge in a other brane contexts), we believe these examples
illustrate the phenomenological interest of string theory
compactifications with branes at singularities.

\end{document}